\documentclass[aps,pra,superscriptaddress,twocolumn]{revtex4}
\usepackage{graphicx,amsmath,amssymb,amsfonts,latexsym,color,dcolumn,bm}

\newcommand{\beq}{\begin{equation}}
\newcommand{\eeq}{\end{equation}}
\newcommand{\bea}{\begin{eqnarray}}
\newcommand{\eea}{\end{eqnarray}}
\providecommand{\abs}[1]{\left\lvert#1\right\rvert}

\providecommand{\bra}[1]{\langle #1 \rvert}
\providecommand{\ket}[1]{\lvert #1 \rangle}

\usepackage{amsfonts,amssymb}
\usepackage{dsfont}
\usepackage{physics}
\usepackage{tocvsec2}

\usepackage{appendix}

\usepackage{tikz}

\usepackage{amsmath}
\usepackage{bbold}
\usepackage{hyperref}
\hypersetup{
     colorlinks=true,      
    linkcolor=blue,        
    citecolor=blue,        
    filecolor=magenta, 
    urlcolor=black          
}

\usepackage{comment}

\begin{document}

\author{N. Fabre\footnote{nfabre@ucm.es}}
\affiliation{Centre for Quantum Optical Technologies, Centre of New Technologies, University of Warsaw, ul.  Banacha 2c, 02-097 Warszawa, Poland
}
\affiliation{Departamento de Óptica, Facultad de Física, Universidad Complutense, 28040 Madrid, Spain}

\date{\today}
\begin{abstract}
In this paper, we investigate the influence of the symmetry of the biphoton wavefunction on the coincidence measurement of the generalized Mach-Zehnder (MZ) interferometer, where there are a temporal and frequency shift operations between the two beam-splitters. We show that the generalized MZ interferometer allows the measurement of the short-time Fourier transform of the function modeling the energy conservation of a spontaneous parametric down-conversion process if the full biphoton state is symmetric, and of the symmetric characteristic distribution of the phase-matching function if the state is antisymmetric. Thus, this technique is phase-sensitive to the spectral distribution of the photon pairs. Finally, we investigate in detail the signature of a pair of anyons whose peculiar statistics can be simulated by engineering the spectrum of photon pairs. 
 \end{abstract}
\vskip2pc 

\title{Interferometric signature of different spectral symmetries of biphoton states}
\maketitle

\section{Introduction}

Quantum information can be carried by many possible degrees of freedom of single photons, which are described by discrete or continuous variables. Among them, the frequency degree of freedom has attracted particular attraction because there are no linear physical processes affecting the spectral distribution of single photons. Besides, the frequency degree of freedom provides a large dimensional encoding which allows reducing the optical resources for all optical-network \cite{joshi_frequency-domain_2020,kues_quantum_2019,kues_-chip_2017}.  Photon pairs can be produced by spontaneous parametric down-conversion or four-wave mixing processes, with bulk, integrated optical devices \cite{autebert_integrated_2016,luo_counter-propagating_2020} and atomic-cooled systems \cite{mazelanik_temporal_2020,PhysRevLett.127.163601}. Photon pairs find direct applications in quantum communication \cite{yin_entanglement-based_2020}, imaging \cite{ndagano_hong-ou-mandel_2021,padgett_introduction_2017}.  In quantum-sensing protocols, the quantum enhancement  of using photon pairs can be modest. Indeed, the metrological gain for the determination of an unknown parameter owing to the entanglement of two photons is equal to two, in quantum optical coherent tomography \cite{PhysRevA.86.023820,teich_variations_2012} for instance. However, it goes beyond that.  Single photons can be used for sensing fragile materials such as biological samples. In addition, single-photon interferometry does not need  phase stabilization in contrast to classical fields to achieve the same performance in metrological tasks. Hong-Ou-Mandel metrology was developed based on these features \cite{chen_hong-ou-mandel_2019,lyons_attosecond-resolution_2018,fabre_parameter_2021,harnchaiwat_tracking_2020,scott_beyond_2020}. Thus, the engineering of the spectral distribution of photon pairs is an interesting line of study for near-term experiments that do not require many entangled single photons.  However, the experimental manipulation of the frequency spectrum of a photon pair  is challenging because it requires either non-linear effects at the single photon level (see for instance \cite{guerreiro_nonlinear_2014,fabre_spectral_2021}) or the use of successive optical elements that drastically decrease the brightness of the source (see for the last point, experiments with weak coherent states in \cite{lukens_frequency-encoded_2017,lu_controlled-not_2019}). Furthermore, for these reasons, methods for performing the spectra-temporal tomography of classical fields can not be applied in the single photon regime (or photon pairs), and other techniques have been investigated in such a regime \cite{jizan_phase-sensitive_2016,thekkadath_measuring_2022,PhysRevLett.127.163601}. \\

Different spectrum symmetries of photon pairs are required in quantum information protocols. For instance, photon pairs with an antisymmetric distribution of degree of freedom have applications  in quantum computation and communication protocols \cite{https://doi.org/10.1002/prop.200310021,goyal_qudit-teleportation_2014}.   Anyons, which are quasiparticles with fractional statistics, have been observed only recently experimentally in quantum electronics experiments \cite{bartolomei_fractional_2020}. They are conjectured to exist only in low dimensions and consist of interacting fermions. The spectrum of photon pairs can be shaped to simulate the particle's statistics of fermions \cite{francesconi_engineering_2020} or anyons \cite{francesconi_anyonic_2021,sansoni_two-particle_2012,jin_quantum_2018}. However, note that in such proposals \cite{francesconi_engineering_2020,francesconi_anyonic_2021,sansoni_two-particle_2012,jin_quantum_2018}, there is no physical process involved creating effective fermions or anyons particles from interacting bosons. Practical devices for simulating different particle's statistics are the following integrated waveguide optical device described in \cite{francesconi_engineering_2020, francesconi_anyonic_2021} and also the atomic-cooled system presented in \cite{zhao_shaping_2015}. Because, in such examples, the spatial transverse profile of the pump corresponds to the phase-matching function of the photon pair.  By engineering the spectrum of photon pairs before their generation, we circumvent the drawback of manipulating the spectrum of single photons by decreasing the total brightness of the field. The signature of the symmetry of the spectral distribution of photon pairs can be obtained using the Hong-Ou-Mandel interferometer \cite{HOM,PhysRevA.56.1627,grice_spectral_1998}.   By fixing the amplitude distribution of the degrees of freedom to be symmetric or separable, other than the spectral one of distinguishable photon pairs, a photon pair with a symmetric spectrum gives rise to a dip in the coincidence measurement, a feature associated with bosonic statistics. While a pair of bosons with antisymmetric spectrum gives rise to an anti-dip, a feature associated with fermionic statistics.  The peculiar statistic of anyons is characterized by a phase accumulated during the exchange of particles, $e^{i\pi\alpha}$ where $\alpha=[0,1]$. The coincidence probability obtained with the HOM interferometer shows that, when a photon pair with an anyonic symmetric is considered, a continuous deformation from the dip obtained with bosonic statistics $\alpha=0$ to the anti-dip obtained with fermionic $\alpha=1$ statistics.\\
 
In this paper, we study the signature of different spectral symmetries of photon pairs created by spontaneous parametric down-conversion (SPDC) revealed by the coincidence probability measured with a Mach-Zehnder (MZ) interferometer.  The different shaping of the spectral distribution of photon pairs allows  the simulation of the peculiar statistics  obtained with fermions or anyons particles. We found that if the joint spectrum amplitude (JSA) is symmetric (resp. antisymmetric) by the exchange of photons, the coincidence probability depends only on the cosine Fourier transform of the function modeling the energy conservation $f_{+}$ of the nonlinear process (resp. the phase-matching function $f_{-}$). However, with the HOM interferometer, the coincidence probability depends always on the phase-matching function of the photon pair regardless of the symmetry of the state \cite{douce_direct_2013,boucher_toolbox_2015}. For an anyonic symmetry, the coincidence probability measured with the HOM interferometry corresponds to the coherent sum between the symmetric and antisymmetric components of the phase-matching function. Whereas for the MZ interferometry, the coincidence probability is the incoherent sum of  the cosine Fourier transform of the energy conservation function and the phase-matching one.

 In addition, we  introduce a novel type of non-linear MZ interferometer, which consists of a frequency beam-splitter (FBS) introduced in \cite{fabre_spectral_2021} placed between two balanced beam-splitters. This nonlinear optical element introduces a non-trivial which-path information that destroys the coherence of the antisymmetric parts but preserves the one of the symmetric part of the full spectrum of the photon pair.
Finally, we introduce the generalized MZ interferometer, consisting of a usual MZ interferometer but with an additional frequency shift between the two balanced beam-splitters. In such a case,  if the full JSA is symmetric (resp. antisymmetric),  the coincidence probability corresponds to the normal (resp. symmetric) characteristic distribution modeling the energy conservation $f_{+}$ (resp. the phase-matching function $f_{-}$) of the SPDC process. Thus, the shaping of the symmetry of the JSA of photon pairs allows measuring, thanks to the generalized MZ interferometer,  either the amplitude and phase of the $f_{+}$ function or  the $f_{-}$ one.\\

The paper is organized as follows. In Sec.~\ref{sectwo}, we study the different coincidence probability outcomes of the  Mach-Zehnder interferometer depending on the different symmetries of the biphoton state and compare them with those obtained using the HOM interferometer. In Sec.~\ref{generalized}, we introduce the generalized Mach-Zehnder interferometry  which allows the measurement of the characteristic distribution of the photon pair. In Sec.~\ref{secthree}, we perform a comparative study obtained with both interferometers of the signature of photon pairs having an anyonic symmetry. Finally, in Sec.~\ref{conclusion}, we summarize our results and provide new perspectives.

\section{Signature of different spectral symmetries of biphoton states with Hong-Ou-Mandel and Mach-Zehnder interferometry}\label{sectwo}
In this section, we start by reminding the influence of the symmetry of the spectral biphoton wavefunction in the expression of the coincidence probability measured with the HOM interferometer. Then, we study the signature of the symmetric and antisymmetric property of the JSA in the coincidence measurement obtained with the MZ interferometer. We finish the section by introducing a new type of nonlinear MZ interferometer.

\subsection{Hong-Ou-Mandel interferometry}\label{homgeneral}
The photon pair produced by a type-II SPDC or a four-wave mixing process can be described by the wavefunction:
\begin{equation}\label{SPDC}
\ket{\psi}=\int_{\mathds{R}}\int_{\mathds{R}} d\omega_{s} d\omega_{i} \text{JSA}(\omega_{s},\omega_{i}) \hat{a}^{\dagger}(\omega_{s})\hat{b}^{\dagger}(\omega_{i}) \ket{0},
\end{equation}
where $a,b$ denote two spatial (or polarizations) modes. The JSA  can generally be decomposed as follows:
\begin{equation}\label{normalfacto}
\text{JSA}(\omega_{s},\omega_{i})=f_{+}(\omega_{+})f_{-}(\omega_{-}),
\end{equation}
where $\omega_{\pm}=(\omega_{s}\pm \omega_{i})/2$, $f_{+}$ is a function that models the energy conservation of the nonlinear process and depends on the spectral profile of the pump. Such a function is a delta distribution in the case of continuous pump degenerate down-conversion, which will not be the case in this study. We consider a wideband pump field, namely ultrashort pump pulses. $f_{-}$ is the phase-matching function which can have various forms depending on the considered nonlinear crystal and the method to achieve phase-matching. The two-photon state then enters the generalized HOM interferometer \cite{douce_direct_2013}, consisting of a time and frequency displacement operation in spatial paths $a$ and $b$, followed by a beam-splitter and non-frequency-resolved coincidence detection. The measured coincidence probability takes the following form \cite{fabre_spectral_2021,douce_direct_2013}:
\begin{equation}\label{wignerhom}
I(\tau,\mu)=\frac{1}{2}(1-\pi W_{-}(\tau,\frac{\mu}{2}))
\end{equation}
where the chronocyclic Wigner distribution of the phase-matching function $f_{-}$ is defined as:
\begin{equation}
W_{-}(\tau,\mu)=\frac{1}{\pi} \int_{\mathds{R}} e^{2i\omega_{-} \tau} f_{-}(\mu-\omega_{-})f^{*}_{-}(\mu+\omega_{-}) d\omega_{-}.
\end{equation}
Note the presence of the factor $\mu/2$ in the chronocyclic Wigner distribution in Eq.~(\ref{wignerhom}) because of the definition of the collective frequency variable $\omega_{-}=(\omega_{s}-\omega_{i})/2$. In other words, the generalized HOM interferometer is a measurement of the displaced parity operator \cite{fabre_generation_2020} of only the phase-matching part of the full JSA. Importantly, the coincidence probability never depends on the function $f_{+}$.
We will now rewrite Eq.~(\ref{wignerhom}) to emphasize its dependance on the symmetry of the biphoton state. The phase-matching function is decomposed into its symmetric $S$ and  antisymmetric components $\tilde{S}$:
\begin{equation}
f_{-}(\omega_{-})=\text{cos}(\frac{\alpha}{2})S(\omega_{-})+\text{sin}(\frac{\alpha}{2})\tilde{S}(\omega_{-}).
\end{equation}
If the state is symmetric (resp. antisymmetric) then $\alpha=0$ (resp. $\alpha=\pi$). For the general case $\alpha\in ]0,\pi[$, the state is said to have an anyonic symmetry.  The coincidence probability can be rewritten as
\begin{multline}\label{wigner}
I(\tau,\mu)=\frac{1}{2}(1-\text{cos}^{2}(\frac{\alpha}{2})W_{S}(\tau,\frac{\mu}{2})-\text{sin}^{2}(\frac{\alpha}{2})W_{\tilde{S}}(\tau,\frac{\mu}{2})\\
-\text{cos}(\frac{\alpha}{2})\text{sin}(\frac{\alpha}{2})(W_{S\tilde{S}}(\tau,\frac{\mu}{2}))+W_{\tilde{S}S}(\tau,\frac{\mu}{2})),
\end{multline}
where we have defined the chronocyclic cross-Wigner distribution:
\begin{equation}
W_{S\tilde{S}}(\tau,\mu)=\frac{1}{\pi} \int_{\mathds{R}} e^{2i\omega_{-} \tau} S(\mu-\omega_{-})\tilde{S}^{*}(\mu+\omega_{-}) d\omega_{-}.
\end{equation}
HOM interferometry reveals an interference effect between the symmetric and antisymmetric parts of the phase-matching function. When the state is antisymmetric namely $\alpha=\pi$, we have that $W_{\tilde{S}}(0,0)=-1$ and thus $I(0,0)=1$. In this case, the two photons are then distinguishable and reproduce the case of a pair of fermions with a symmetric spectrum. The fermions cannot occupy the same mode, and are described by the fermionic operators described in Appendix \ref{MZfermions}, whose particular algebra is reproduced by the antisymmetry of the phase-matching function. The coincidence probability $I(\tau)$ reveals an anti-dip characteristic of anti-bunching particles. However, for a symmetric spectrum, namely $\alpha=0$, since we have  $W_{S}(0,0)=1$, we deduce that $I(0,0)=0$. Thus, no coincidence counts are expected from the symmetric part which is interpreted by two indistinguishable photons bunch.   Another equivalent writing is by writing the phase-matching function as $f_{-}(\omega_{-})=e^{i\alpha}f^{*}_{-}(-\omega_{-})$, we obtain directly the expression of the coincidence $I(0,0)=1/2(1-\text{cos}(\alpha))$. Interestingly, for $\alpha=\pi/2$, the photons are distinguishable while no temporal delay is introduced, interpreted as an interference effect between the symmetric and antisymmetric part of the JSA. More details regarding the anyonic case are presented in Sec.~\ref{secthree}. This new writing clarifies why the HOM experiment is said to be a symmetric filter \cite{PhysRevA.94.033855,fedrizzi_anti-symmetrization_2009}, since the antisymmetric part remains. The MZ interferometer is not a symmetric filter of the degree of freedom distribution of photon pairs, but also captures the interference between the symmetric and antisymmetric parts of the \textit{full} JSA, as shown in the next section.

\subsection{Mach-Zehnder interferometry}
\begin{figure*}
 \begin{center}
\includegraphics[width=1\textwidth]{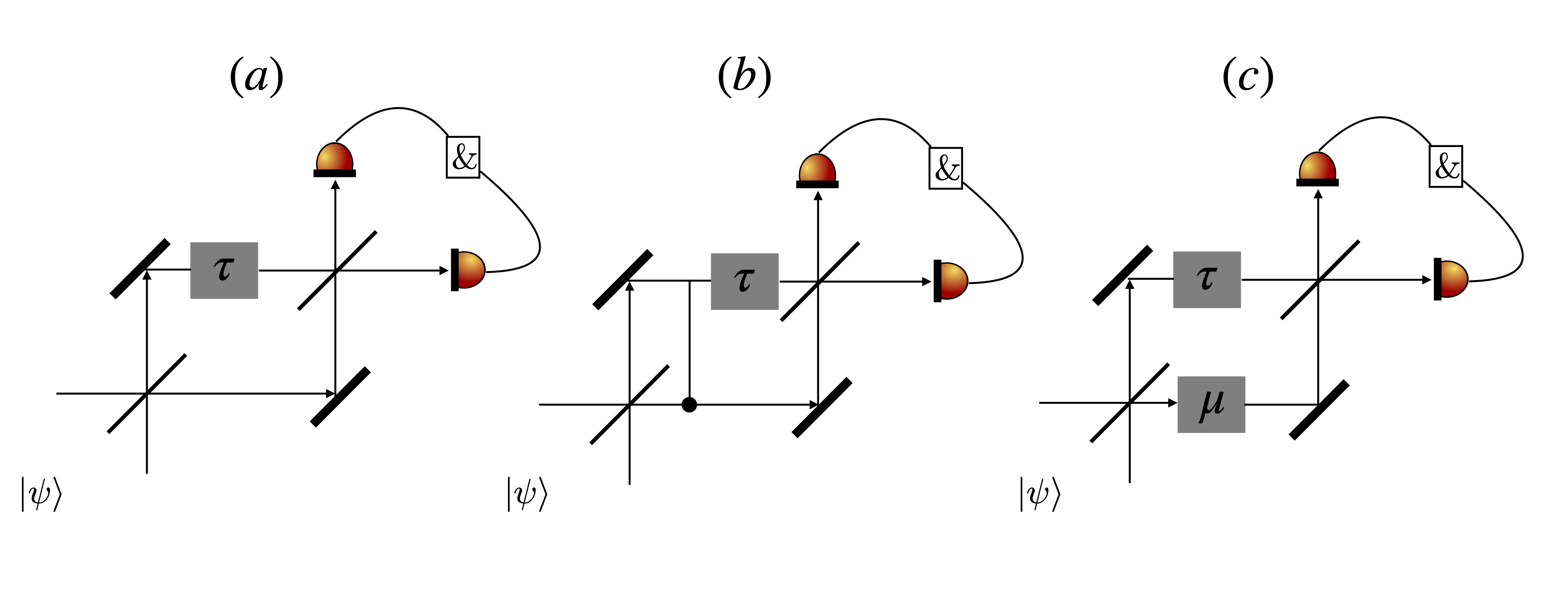}
\caption{\label{Machzender}(a) Mach-Zehnder interferometer, consisting of two-balanced beam-splitters, a time displacement operation inside them, and a coincidence measurement. (b) Nonlinear Mach-Zehnder interferometer. A frequency beam-splitter is placed between the two arms inside the interferometer, which is represented as a CNOT operation. (c) Generalized Mach-Zehnder interferometer. A frequency and temporal shift operations are performed between the two balanced beam-splitters. }
\end{center}
\end{figure*}

The biphoton state described by Eq.~(\ref{SPDC}) is now the input to the MZ interferometer as shown in Fig.~\ref{Machzender}(a). This interferometer is composed of a first balanced beam-splitter, followed by a time displacement operation in one of the spatial arms, then a second balanced beam-splitter ended by a non-resolved frequency coincidence measurement.  The non-frequency-resolved coincidence probability can be expressed as
\begin{multline}\label{difficultexpression}
I(\tau)=\frac{1}{16}\int_{\mathds{R}}\int_{\mathds{R}} |(\text{JSA}(\omega_{s},\omega_{i}) +\text{JSA}(\omega_{i},\omega_{s}))(e^{i(\omega_{s}+\omega_{i})\tau}+1) \\
+(e^{i\omega_{s}\tau}+e^{i\omega_{i}\tau})(\text{JSA}(\omega_{s},\omega_{i}) -\text{JSA}(\omega_{i},\omega_{s}))|^{2}d\omega_{s}d\omega_{i} 
\end{multline}
and is presented in Appendix \ref{interferometry}. Contrary to the coincidence probability measured with the HOM interferometer, this expression also depends on the function $f_{+}$ owing to the phase term $e^{i\omega_{+}\tau}$.  We now consider specific symmetry of the JSA to simplify and obtain analytical expressions of the coincidence probability.

If the JSA is symmetric $\text{JSA}(\omega_{s},\omega_{i})=\text{JSA}(\omega_{i},\omega_{s})$, then the coincidence probability in Eq.~(\ref{difficultexpression}) is reduced to
\begin{equation}\label{symmetryJSA}
I(\tau)=\frac{1}{2}(1+\int_{\mathds{R}}\int_{\mathds{R}} d\omega_{s}d\omega_{i} \text{JSI}(\omega_{s},\omega_{i}) \text{cos}(\omega_{+}\tau)),
\end{equation}
where the JSI stands for joint spectral intensity and is the absolute value square of the JSA. Now, let us consider the factorization of the JSA described by Eq.~(\ref{normalfacto}). The coincidence probability takes the following form:
\begin{equation}\label{symmetriccase}
I(\tau)=\frac{1}{2}(1+\int_{\mathds{R}} d\omega_{+} \abs{f_{+}(\omega_{+})}^{2} \text{cos}(\omega_{+}\tau))=\frac{1}{2}(1+P_{+}(\tau)),
\end{equation}
where $P_{+}(\tau)$ is the cosine Fourier transform of $\abs{f_{+}}^{2}$, which allows us to obtain the spectral width of the pump  \cite{jin_extended_2018}. Indeed, the inverse formula yields \cite{zielnicki_joint_2018}
\begin{equation}
 \abs{f_{+}(\omega_{+})}^{2}\propto \int_{\mathds{R}} d\tau (I(\tau)-\frac{1}{2}I(\tau=0)) \text{cos}(\omega_{+}\tau).
\end{equation}
 The expression Eq.~(\ref{symmetriccase}) explicitly shows that the function $f_{-}$ does not intervene.  In HOM interferometry, the coincidence probability always depends on $f_{-}$ for any symmetry of the JSA. The dependence of the coincidence probability in the function $f_{+}$ lays potential application in shift-estimation protocols. In contrast to the HOM case where only the phase-matching part must be optimized to increase the Fisher information  to reduce the variance of a given parameter to estimate \cite{chen_hong-ou-mandel_2019,lyons_attosecond-resolution_2018,fabre_parameter_2021}, here it is the spectral pump profile that must be optimized for such sensing applications.\\

 If the JSA is antisymmetric, $\text{JSA}(\omega_{s},\omega_{i})=-\text{JSA}(\omega_{i},\omega_{s})$, then the coincidence probability  in Eq.~(\ref{difficultexpression})  yields
\begin{equation}
I(\tau)=\frac{1}{2}(1+\int_{\mathds{R}}\int_{\mathds{R}} d\omega_{s}d\omega_{i} \text{JSI}(\omega_{s},\omega_{i}) \text{cos}(\omega_{-}\tau)).
\end{equation}
By assuming again the factorization Eq.~(\ref{normalfacto}), we obtain:
\begin{equation}\label{fermionic}
I(\tau)=\frac{1}{2}(1+\int_{\mathds{R}} d\omega_{-} \abs{f_{-}(\omega_{-})}^{2} \text{cos}(\omega_{-}\tau))=\frac{1}{2}(1+P_{-}(\tau)),
\end{equation}
where $P_{-}(\tau)$ denotes the cosine Fourier transform of $\abs{f_{-}}^{2}$. In this case,  the function $f_{+}$ does not intervene. For the two considered symmetries, the coincidence is composed of two terms: the average number of single counts and an interference term whose the oscillation period is $2\pi/\omega_{c}^{\pm}$ ($\omega_{c}^{\pm}$ being the central frequency of the functions $f_{\pm}$). In conclusion, the signature of the spectral symmetry of the photon pair with the MZ interferometer is the dependence of the coincidence probability on the function depending on the collective variable $\omega_{+}$ (resp. $\omega_{-}$) of the full JSA, if the state is symmetric (resp. antisymmetric). Note that the visibility of the oscillation is reduced when the purity of the state decreases.  In Appendix \ref{MZfermions}, we show that the coincidence probability obtained with the MZ interferometer starting from a pair of fermions with a symmetric spectrum reproduces the statistics of a pair of bosons with an antisymmetric spectrum, which is also the case in the HOM experiment.

Note that a single photon described by the wavefunction $\ket{\psi}=\int d\omega S(\omega) \ket{\omega}_{a}$ gives rise to a single-count probability with a mathematical form similar to the coincidence probability  obtained with a photon pair with a symmetric or antisymmetric spectrum. Indeed, the single-count probability measured in the spatial port $a$ or $b$ is
\begin{equation}
P_{a,b}(\tau)=\frac{1}{2}(1\pm \int_{\mathds{R}} \abs{S(\omega)}^{2}\text{cos}(\omega\tau) d\omega).
\end{equation}
Analogous facts have already been observed in HOM interferometry and in one and two-photon Young's experiments \cite{fabre_producing_2020}. Finally, we note that the Wiener-Khinchin theorem has been used to express the expression of the coincidence probability in terms of the first-order temporal coherence of the input light, either with single photon states or with the symmetric and antisymmetric part of the spectrum of photon pair \cite{jin_extended_2018,luo_quantum_2021}. \\

\begin{figure}[h!]
 \begin{center}
\includegraphics[width=0.5\textwidth]{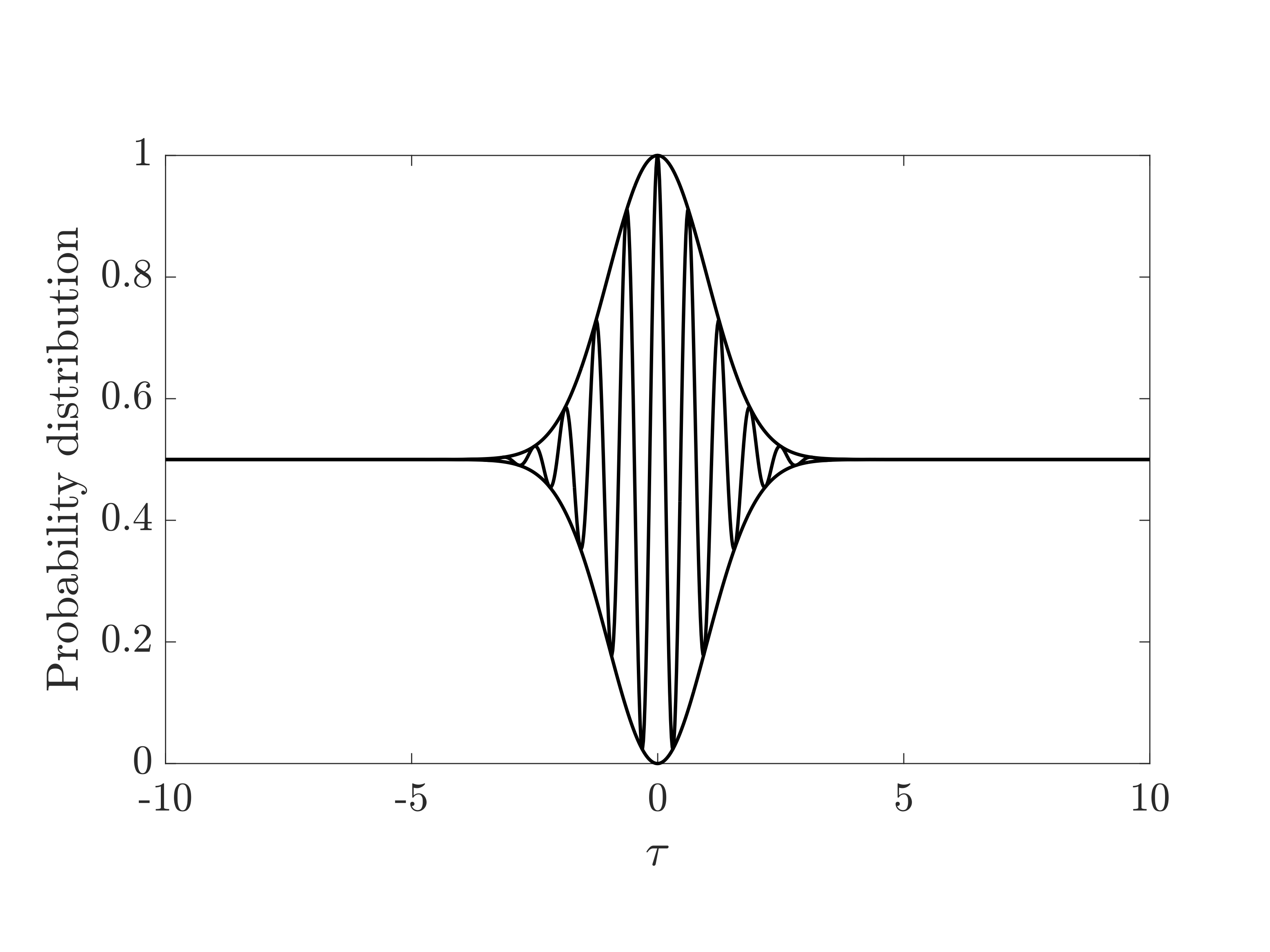}
\caption{\label{scheme} An identical probability distribution with respect to the delay $\tau$  measured with the Mach-Zehnder interferometer corresponding to three different inputs noted (1), (2) and (3). (1) Coincidence probability obtained with photon pairs possessing a symmetric JSA, whose oscillation period depends on the central frequency of $f_{+}$. (2) Coincidence probability obtained with photon pairs possessing an antisymmetric JSA, whose oscillation period depends on the central frequency of $f_{-}$. (3) Single count of one particle MZ, whose oscillation period depends on the central frequency of the spectrum $F$ (with twice the base line). The unit of $\tau$ depends on the photon pairs source, and is typically the picosecond.}
\end{center}
\end{figure}

In Fig.~\ref{scheme}, we represent three different cases that give the same probability distribution: the coincidence probability for a symmetric and antisymmetric JSA of a photon pair and the single-count probability for a single photon. The function $f_{\pm}$ and $S$ has been chosen Gaussian while the antisymmetric JSA is under the form $f_{-}(\omega_{-})=\text{sgn}(\omega_{-})e^{-\omega_{-}^{2}/\sigma^{2}}$. This choice corresponds to the most simple antisymmetric function close to the Gaussian one. In other words, Fig.~\ref{scheme} emphasizes that the same probability distribution can be obtained from different inputs, either with single photons or with different spectral entanglement of photon pairs.\\

\subsection{Effect of frequency beam-splitter operation placed inside a Mach-Zehnder interferometer}

We propose a method to suppress the coherence of the antisymmetric part of the spectral biphoton state by placing a non-linear optical element called a frequency beam-splitter (FBS) \cite{fabre_spectral_2021} between the two balanced beam-splitters of a MZ interferometer.  The FBS performs the following operation $\hat{U}\ket{\omega_{s},\omega_{i}}_{ab}=\ket{\omega_{+},\omega_{-}}_{ab}$ and its general action on a biphoton state is the following:
\begin{multline}
\hat{U}\ket{\psi}=\int_{\mathds{R}}\int_{\mathds{R}} d\omega_{s} d\omega_{i} \text{JSA}(\omega_{s},\omega_{i}) \ket{\omega_{+},\omega_{-}}_{ab}\\
= \int_{\mathds{R}}\int_{\mathds{R}} d\omega_{s} d\omega_{i} \text{JSA}(\omega_{+},\omega_{-}) \ket{\omega_{s},\omega_{i}}_{ab}.
\end{multline}
Other mathematical details  of the FBS can be found in Appendix \ref{Frequencybeamsp}. In \cite{fabre_spectral_2021}, we propose a direct experimental implementation of the FBS using cascade non-linear effect. Such an operation could be performed with split-ring resonators  \cite{joshi_frequency-domain_2020,lu_periodically_2019,yang_squeezed_2021,grassani_micrometer-scale_2015,kues_quantum_2019} or atoms coupled to two waveguides \cite{oehri_tunable_2015}. We assume that the gate does not perform any operation if two photons  enter one spatial port $a$ or $b$ of the FBS, which is a supposed characteristic of the device implementing such an operation.

The state produced by the SPDC process described in Eq.~(\ref{SPDC}) enters into the interferometer represented in Fig.~\ref{Machzender}(b), called the non-linear MZ interferometer. The interferometer is composed of a time displacement operation and a frequency beam-splitter operation between two balanced beam-splitters. After the first balanced beam-splitter, the FBS and the time displacement operation, the wavefunction of the biphoton state can be written as
\begin{multline}
\ket{\psi_{\tau}}=\frac{1}{2}\int_{\mathds{R}}\int_{\mathds{R}} d\omega_{s}d\omega_{i}[\text{JSA}(\omega_{s},\omega_{i})e^{i\omega_{+}\tau} \hat{a}^{\dagger}(\omega_{s})\hat{a}^{\dagger}(\omega_{i})\\
- \text{JSA}(\omega_{+},\omega_{-})e^{i\omega_{+}\tau}\hat{a}^{\dagger}(\omega_{s})\hat{b}^{\dagger}(\omega_{i})\\
+\text{JSA}(\omega_{-},\omega_{+})e^{i\omega_{+}\tau}\hat{a}^{\dagger}(\omega_{s})\hat{b}^{\dagger}(\omega_{i})\\
-\text{JSA}(\omega_{s},\omega_{i})\hat{b}^{\dagger}(\omega_{s})\hat{b}^{\dagger}(\omega_{i})]\ket{0}.
\end{multline}
 The post-selected wavefunction that gives rise to coincidence events after the second beam-splitter is
\begin{multline}
\ket{\psi_{\tau}}=\frac{1}{4}\int_{\mathds{R}}\int_{\mathds{R}} [(\text{JSA}(\omega_{s},\omega_{i})+\text{JSA}(\omega_{i},\omega_{s}))(e^{i\omega_{+}\tau}+1)\\
+e^{i\omega_{+}\tau}(\text{JSA}(\omega_{-},\omega_{+})-\text{JSA}(-\omega_{-},\omega_{+})\\
- \text{JSA}(\omega_{+},\omega_{-})+\text{JSA}(\omega_{+},-\omega_{-})) \ket{\omega_{s},\omega_{i}}_{ab}d\omega_{s}d\omega_{i}.
\end{multline}
 If the JSA is symmetric, we obtain the same result without the FBS (see Eq.~(\ref{symmetryJSA})). We have no more a phase term that depends on $\omega_{-}$, which is required to observe the interference of the antisymmetric part. If the state is antisymmetric, the wavefunction can be expressed as
\begin{multline}
\ket{\psi_{\tau}}=\frac{1}{2}\int_{\mathds{R}}\int_{\mathds{R}} d\omega_{s}d\omega_{i} (\text{JSA}(\omega_{-},\omega_{+})+\text{JSA}(\omega_{+},-\omega_{-}))\\
\cross e^{i\omega_{+}\tau}\ket{\omega_{s},\omega_{i}}_{ab}.
\end{multline}
Assuming that the decomposition of the JSA described by Eq.~(\ref{normalfacto}), and assuming that $f_{+}$ is a symmetric function, the coincidence probability can be written as
\begin{equation}
I(\tau)=\frac{1}{2}(1-\int_{\mathds{R}}\int_{\mathds{R}} d\omega_{+}d\omega_{-} \abs{f_{+}(\omega_{+}-\omega_{-})}^{2} \abs{f_{-}(\omega_{+}+\omega_{-})}^{2}).
\end{equation}
The coincidence probability is constant and thus no oscillation pattern is observed. The physical interpretation of this result is that the FBS inside the MZ interferometer adds a which-path information, which prevents revealing the coherence of the antisymmetric part of the biphoton wavefunction.

\section{Generalized Mach-Zehnder interferometry}\label{generalized}
In this section, we introduce the generalized MZ interferometry represented in Fig.~\ref{Machzender}(c), which is a phase-sensitive measurement of the spectral distribution of photon pairs. A frequency displacement operation is placed in the spatial port $b$ as well as the usual time displacement in the spatial port $a$ between the two beam-splitters. Experimentally, a frequency displacement operation can be done with electro-optic modulator or with the device presented in  \cite{hu_-chip_2021}. The goal is to analyze the expression of the coincidence probability as a function of the time and frequency shifts for different symmetries of the JSA. 

\begin{figure*}
 \begin{center}
\includegraphics[width=1\textwidth]{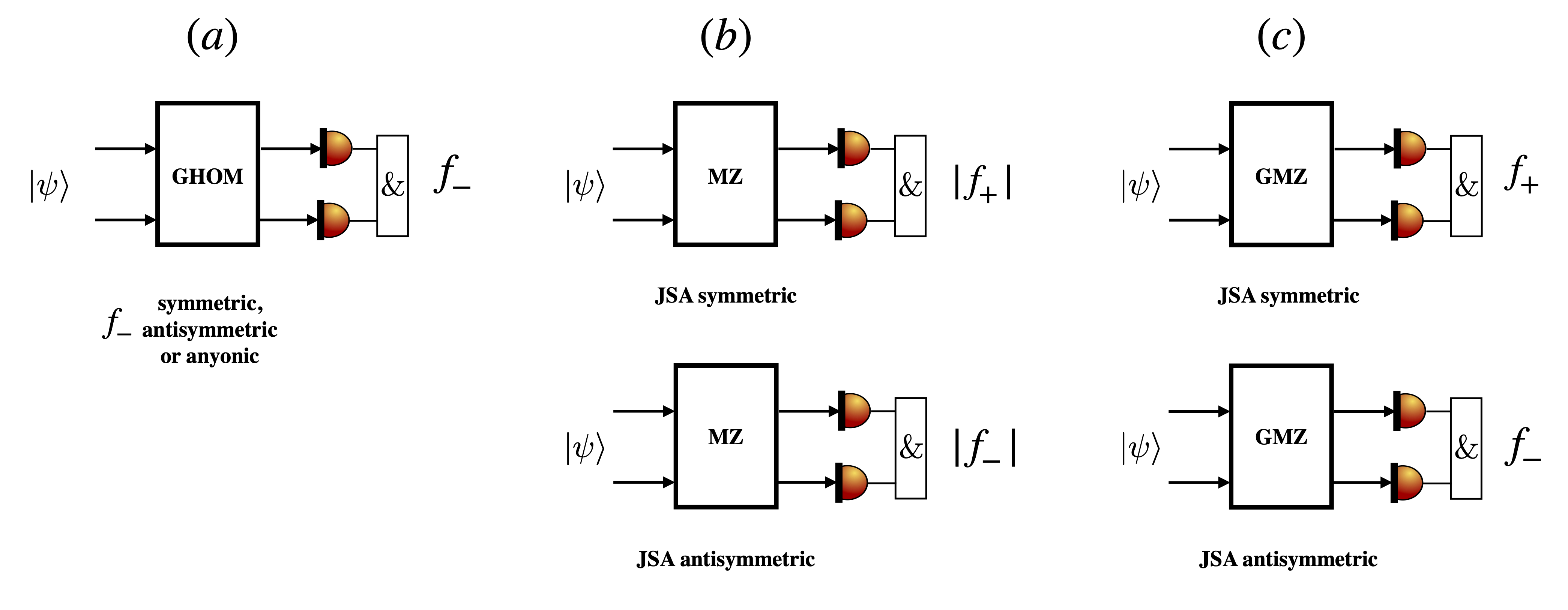}
\caption{\label{recap}(a) Generalized Hong-Ou-Mandel interferometer (GHOM), allowing the measurement of  $f_{-}$ for any symmetry of $f_{-}$. (b) Mach-Zehnder (MZ) interferometer which allows the measurement of the amplitude of $f_{\pm}$, namely $\abs{f_{\pm}}$ if the full biphoton state is symmetric (resp. antisymmetric). (c) Generalized Mach-Zehnder (GMZ) interferometer allowing the measurement of the phase and amplitude of $f_{\pm}$ is the full biphoton state is symmetric (resp. antisymmetric). \& means coincidence measurement. }
\end{center}
\end{figure*}

The post-selected wavefunction leading to coincidence events after the second beam-splitter can be calculated using  calculations similar to those described in the previous section:
\begin{multline}\label{generalwavefii}
\ket{\psi_{\tau,\mu}}=\frac{1}{4}\int_{\mathds{R}}\int_{\mathds{R}} (\text{JSA}(\omega_{s},\omega_{i})+\text{JSA}(\omega_{i},\omega_{s}))e^{i(\omega_{s}+\omega_{i})\tau}\\
+\text{JSA}(\omega_{s},\omega_{i}+\mu)-\text{JSA}(\omega_{i}+\mu,\omega_{s})e^{i\omega_{s}\tau}\\
+\text{JSA}(\omega_{s}+\mu,\omega_{i})-\text{JSA}(\omega_{i},\omega_{s}+\mu)e^{i\omega_{i}\tau}\\
+(\text{JSA}(\omega_{s}+\mu,\omega_{i}+\mu)+\text{JSA}(\omega_{i}+\mu,\omega_{s}+\mu))\ket{\omega_{s},\omega_{i}}_{ab}  d\omega_{s}d\omega_{i}.
\end{multline}
The first and last terms come from bunching events after the first beam-splitter, while the second and third come from a single photon at each spatial port after the first beam-splitter. We now consider different symmetries of the JSA to simplify the expression of the coincidence probability. If the JSA is symmetric, then the wavefunction reduces to:
\begin{multline}
\ket{\psi_{\tau,\mu}}=\frac{1}{2}\int_{\mathds{R}}\int_{\mathds{R}} d\omega_{s}d\omega_{i} (\text{JSA}(\omega_{s},\omega_{i})e^{i(\omega_{s}+\omega_{i})\tau}\\+\text{JSA}(\omega_{s}+\mu,\omega_{i}+\mu)) \ket{\omega_{s},\omega_{i}}_{ab}.
\end{multline}
In such a case, the only terms that intervene come from two-photon events that cross together the time or the frequency shift between the two beam-splitters. By replacing the factorization of the JSA (see Eq.~(\ref{normalfacto})) in the expression of the coincidence probability, we obtain
\begin{multline}
I(\tau,\mu)=\frac{1}{2}(1+\text{Re}(\int_{\mathds{R}} d\omega_{+} f_{+}(\omega_{+})f^{*}_{+}(\omega_{+}+\mu)e^{2i\omega_{+}\tau}))\\
=\frac{1}{2}[1+\text{Re}(F_{+}(\mu,2\tau))]
\end{multline}
where we define the  short-time Fourier transform (STFT) of the function $f_{+}$ as
\begin{equation}
F_{+}(\mu,2\tau)=\int_{\mathds{R}} d\omega_{+} f_{+}(\omega_{+})f^{*}_{+}(\omega_{+}+\mu)e^{2i\omega_{+}\tau}.
\end{equation}
In general, the STFT of a function is composed of another window function used to analyze the function of interest.  Here, the window function is identical to the function to scan. The absolute square of the STFT is called the spectrogram. The rectangular paving of the time-frequency phase space has one interesting consequence: if the spectral resolution is low, the temporal resolution is high (and vice-versa).  Note that if one introduces a phase $e^{i\pi/2}$ in one of the arms inside the MZ interferometer, then the imaginary part of the STFT is measured. Once the real and imaginary parts are obtained, assuming that $f_{+}(0)\neq 0$, the following reconstruction formula can be used
\begin{equation}
f^{*}_{+}(\mu)=\frac{1}{f_{+}(0)} \int_{\mathds{R}} F_{+}(\mu,\tau) d\tau,
\end{equation}
to perform the full tomography of $f_{+}$. This reconstruction method is similar to the filter bank summation method.  Note that while the STFT is complex, the chronocyclic Wigner distribution is a real distribution. We recover the previous case when $\mu=0$, because the real part of the Fourier transform of $f_{+}$ is the cosine Fourier transform. Now, if the JSA is antisymmetric, the wavefunction in Eq.~(\ref{generalwavefii}) is reduced to:
\begin{multline}
\ket{\psi}=\frac{1}{2}\int_{\mathds{R}}\int_{\mathds{R}} d\omega_{s}d\omega_{i} (\text{JSA}(\omega_{s},\omega_{i}+\mu)e^{i\omega_{s}\tau}\\
-\text{JSA}(\omega_{s}+\mu,\omega_{i})e^{i\omega_{i}\tau})\ket{\omega_{s},\omega_{i}}
\end{multline}
and the coincidence probability can be expressed as
\begin{multline}
I(\tau,\mu)=\frac{1}{2}(1+\text{Re}(\int_{\mathds{R}} d\omega_{-} f_{-}(\omega_{-}-\frac{\mu}{2})f_{-}^{*}(\omega_{-}+\frac{\mu}{2})e^{2i\omega_{-}\tau}))\\
=\frac{1}{2}(1+\text{Re}(e^{i\mu\tau}F_{-}(\mu,2\tau))).
\end{multline}
The expression for the coincidence probability is different from the case in which the JSA is symmetric because of this additional phase factor in the real part $e^{i\mu\tau}$. We find the symmetric characteristic distribution of the function $f_{-}$, namely the average value of the symmetric time-frequency displacement operator $\hat{D}(\mu,\tau)$:
\begin{multline}
\chi_{\psi}(\mu,\tau)=\bra{\psi}\hat{D}(\mu,\tau)\ket{\psi}\\=
e^{i\mu\tau/2}\int_{\mathds{R}} f_{-}(\omega_{-}-\mu)f_{-}^{*}(\omega_{-})e^{i\omega_{-}\tau}d\omega_{-}
\end{multline}
 where $\hat{D}(\mu,\tau)=e^{i\mu\tau/2} \int d\omega e^{i\omega\tau} \ket{\omega+\mu}\bra{\omega}$ (more details can be found in \cite{fabre_generation_2020} about this operator) and we formally define the state $\ket{\psi}=\int f_{-}(\omega_{-}) \ket{\omega_{-}} d\omega_{-}$, because the spectral part depending on $\omega_{+}$ does not have any importance. From this definition, we deduce that the generalized MZ interferometer when the entangled spectrum of the photon pair is symmetric is the measurement of  the STFT which corresponds to the normal characteristic distribution of $f_{+}$ because the normal order of time and frequency displacement operator implies the absence of the phase $e^{i\mu\tau/2}$. We now investigate the case of two separable single photons $\text{JSA}(\omega_{s},\omega_{i})=\gamma(\omega_{s})\beta(\omega_{i})$, the coincidence probability can be cast under the form:
\begin{equation}
I(\tau,\mu)=\frac{1}{2}(1+\text{Re}(F_{\gamma}(\tau,\mu)F_{\beta}(\tau,\mu))).
\end{equation}
We obtain the real part product of the spectrogram of  the spectral function of each single photon state entering into the interferometer.
In Fig.~\ref{recap}, we recap the different HOM and MZ interferometers configurations and what can be measured in each case.

\section{Simulation of anyons statistics with photon pairs}\label{secthree}
In this section, we explicit two definitions of anyons with many degrees of freedom. We will lay the emphasis on one definition and how their statistics can be reproduced with photon pairs. We investigate the signature of the anyonic symmetry with the HOM and MZ interferometers.

\subsection{Definitions of anyons possessing one discrete and one continuous degree of freedom}
As explained in Sec.~\ref{homgeneral}, the HOM interferometry is a direct measurement of the chronocyclic Wigner distribution of the phase-matching function of photon pairs produced by a SPDC process. It is possible to simulate various particle statistics by engineering the spectrum of photons pairs, and in particular the anyonic one.
Multicomponent anyons as defined in \cite{doi:10.1142/S0129055X20300046} are anyons with many degrees of freedom. We consider such anyons possessing a spatial (discrete) and frequency (continuous) degree of freedom. The relation between different creation anyonic operators at spatial path noted 1,2 and with different frequencies is 
\begin{equation}\label{multi}
\hat{a}_{1}^{\dagger}(\omega_{s})\hat{a}_{2}^{\dagger}(\omega_{i})=A(\omega_{i},\omega_{s})\hat{a}_{1}^{\dagger}(\omega_{i})\hat{a}_{2}^{\dagger}(\omega_{s})
\end{equation}
while the relation between different annihilation operators is
\begin{equation}
\hat{a}_{1}(\omega_{s})\hat{a}_{2}(\omega_{i})=A(\omega_{i},\omega_{s})\hat{a}_{1}(\omega_{i})\hat{a}_{2}(\omega_{s})
\end{equation}
and we also have
\begin{equation}
\hat{a}_{1}(\omega_{s})\hat{a}^{\dagger}_{2}(\omega_{i})-A(\omega_{i},\omega_{s})\hat{a}_{1}(\omega_{i})\hat{a}^{\dagger}_{2}(\omega_{s})=\delta(\omega_{s}-\omega_{i})\mathds{I}.
\end{equation}
The function $A$ verifies the properties:
\begin{align}\label{multicomponent}
A(\omega_{s},\omega_{i})=A^{*}(\omega_{i},\omega_{s})\\
\abs{A(\omega_{s},\omega_{i})}=&1
\end{align}
The conjugation operation is fundamental to achieve anyonic symmetry, which is not possible only with permutations. For example, a possible form for $A$ is 
\begin{equation}
A(\omega_{s},\omega_{i})=e^{i\alpha\pi \text{sgn}(\omega_{s}-\omega_{i})}= \left\{
    \begin{array}{ll}
        e^{i\alpha\pi} & \omega_{s}>\omega_{i}\\
       e^{-i\alpha\pi}& \omega_{i}>\omega_{s}
    \end{array}
\right.
\end{equation}
where $\alpha\in [0,1]$. The particular phase accumulation characteristic of the anyonic symmetry when exchanging the particles is $A(\omega_{s},\omega_{i})=e^{2i\pi\alpha} A(\omega_{i},\omega_{s})$. \\

It is also possible to define multicomponent anyons but with the property $\abs{A(\omega_{s},\omega_{i})}<1$  and the corresponding example is \cite{doi:10.1142/S0129055X20300046} :
\begin{equation}
A(\omega_{s},\omega_{i})=ke^{i\alpha \text{sgn}(\omega_{s}-\omega_{i})}= \left\{
    \begin{array}{ll}
        ke^{i\alpha\pi} & \omega_{s}>\omega_{i}\\
       ke^{-i\alpha\pi}& \omega_{i}>\omega_{s}
    \end{array}
\right.
\end{equation}
with $k\in [-1,1]$. However, the examples of multicomponent anyons presented here do not extrapolate with the fermionic statistics. Indeed, for  $\alpha=1$, the function $A$ is not antisymmetric and is equal to $-1$ for any  sign of $\omega_{-}$.\\

We now provide examples of spectral distributions of anyons that allow the extrapolation with the fermionic case when $\alpha=1$. In the examples that we will provide, the function $A$ does not verify the property: $A(\omega_{s},\omega_{i})=A^{*}(\omega_{i},\omega_{s})$, nevertheless,  the relation $A(\omega_{s},\omega_{i})=e^{i\phi} A(\omega_{i},\omega_{s})$ will remain. This last relation allows keeping the peculiar accumulation of the phase during the exchange of particles. As a first example of what we call the multicomponent anyons that extrapolate with the fermionic statistics, we propose
\begin{equation}
A(\omega_{s},\omega_{i})=\text{sgn}(\omega_{s}-\omega_{i})^{\alpha}= \left\{
    \begin{array}{ll}
        1 & \omega_{s}>\omega_{i}\\
       (-1)^{\alpha} & \omega_{i}>\omega_{s}
    \end{array}
\right.
\end{equation}
and the second example is
\begin{equation}
A(\omega_{s},\omega_{i})=(\omega_{s}-\omega_{i})^{\alpha}= \left\{
    \begin{array}{ll}
        (\omega_{s}-\omega_{i})^{\alpha} & \omega_{s}>\omega_{i}\\
       (\omega_{s}-\omega_{i})^{\alpha} (-1)^{\alpha} & \omega_{i}>\omega_{s}.
    \end{array}
\right.
\end{equation}

Now, the bosonic commutation relation traducing the frequency exchange of photons in different spatial paths is $\hat{a}^{\dagger}_{1}(\omega_{s})\hat{a}^{\dagger}_{2}(\omega_{i})=\hat{a}^{\dagger}_{1}(\omega_{i})\hat{a}^{\dagger}_{2}(\omega_{s})$ and does not give rise to the creation of such a function $A$.  However, the phase-matching function of photon pairs can be engineered to create this $A$ function. Then, the full wavefunction composed of the spectrum and the bosonic operators will be equivalent of a pair of anyons with a symmetric spectrum. Again, we do not consider a physical mechanism of interacting bosons system that creates effective anyons. Only the spectrum of photons is shaped to reproduce the anyonic phase.   The first example of a phase-matching function that can be considered is
\begin{equation}\label{firstcasespec}
f_{-}(\omega_{-})=\text{sgn}(\omega_{-})^{\alpha}e^{-\frac{\omega_{-}^{2}}{2\sigma^{2}}}=\left\{
    \begin{array}{ll}
    e^{-\frac{\omega_{-}^{2}}{2\sigma^{2}}} & \omega_{s}>\omega_{i}\\
       (-1)^{\alpha} e^{-\frac{\omega_{-}^{2}}{2\sigma^{2}}} & \omega_{i}>\omega_{s},
    \end{array}
\right.
\end{equation}
where $\text{sgn}(\omega_{-})^{\alpha}$ is the function previously called $A$, and the Gaussian part of $f_{-}$ ensures the normalization of the wavefunction. Another example is the ground state of a pair of anyons. Such a spatial wavefunction can be decomposed into two parts, one depending on the center of mass and the other on the relative motion depending on the difference in the spatial coordinates of the two anyons \cite{PhysRevD.46.2714}. The spatial pump profile simulates the center of mass part, whereas the part that depends on the relative motion is reproduced in the spectral domain with the phase-matching profile of a SPDC process:
\begin{equation}\label{anyonsspec}
f_{-}(\omega_{-})=\omega_{-}^{\alpha}e^{-\frac{\omega_{-}^{2}}{2\sigma^{2}}}
\end{equation}
and can be considered as a special case of the Laughlin state \cite{PhysRevLett.50.1395}. This phase-matching function Eq.~(\ref{anyonsspec}) can be engineered in optical devices where the phase-matching function depends on the collective variable $\omega_{-}$ and is directly proportional to the spatial profile of the pump, in integrated optical circuit \cite{francesconi_anyonic_2021} or atomic-cooled system \cite{zhao_shaping_2015}.  We have that $f^{*}_{-}(-\omega_{-})=e^{-i\pi\alpha} (\omega_{-}^{\alpha})^{*}e^{-\omega_{-}^{2}/2\sigma^{2}}$ and thus such a function does not verify the property of multicomponent anyons Eq.~(\ref{multicomponent}). Nevertheless, the Eq.~(\ref{anyonsspec}) is antisymmetric for $\alpha=1$.

\subsection{Hong-Ou-Mandel interferometry with photon pairs having an anyonic spectrum}
\begin{figure*}
 \begin{center}
\includegraphics[width=1\textwidth]{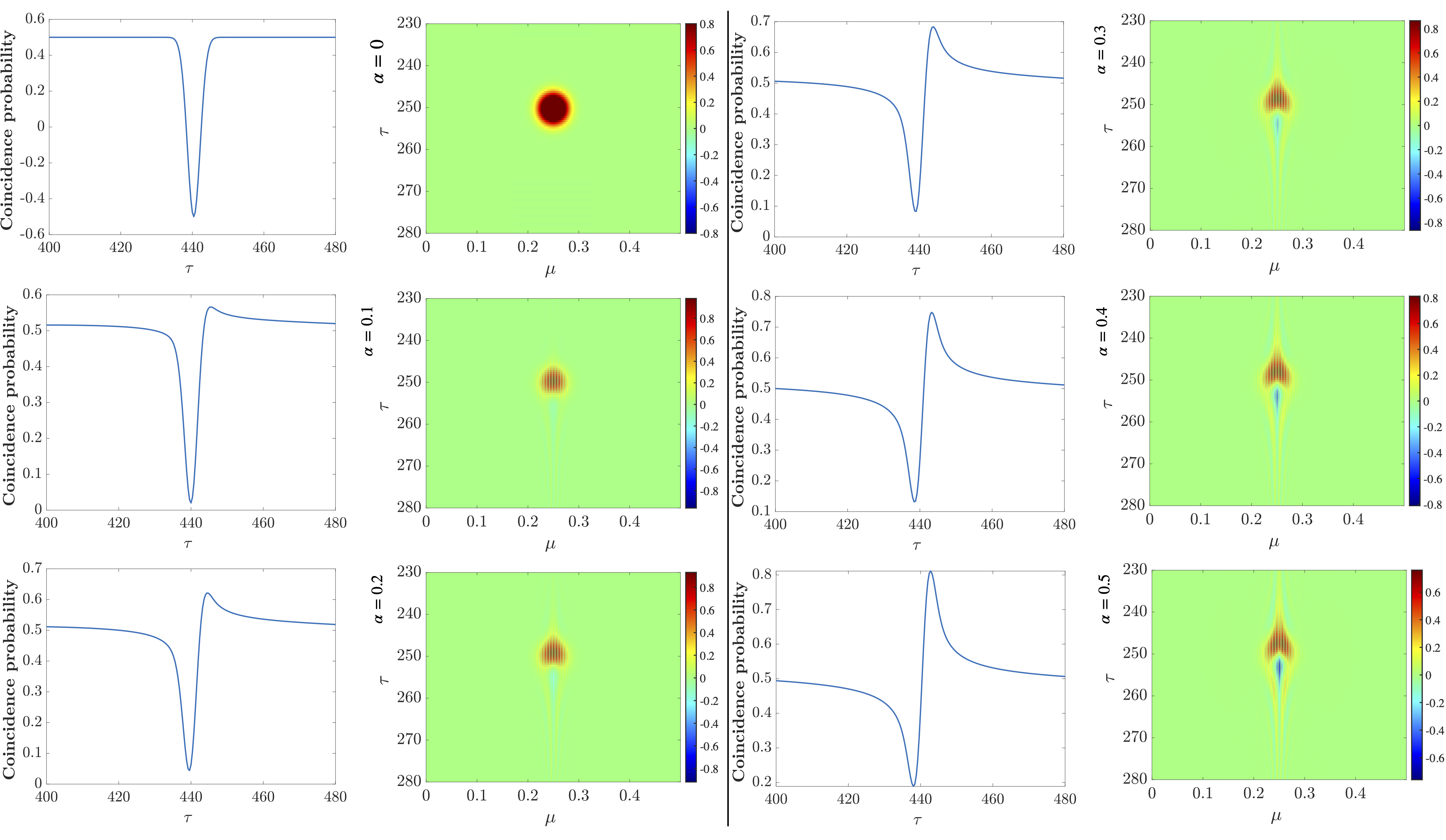}
\includegraphics[width=1\textwidth]{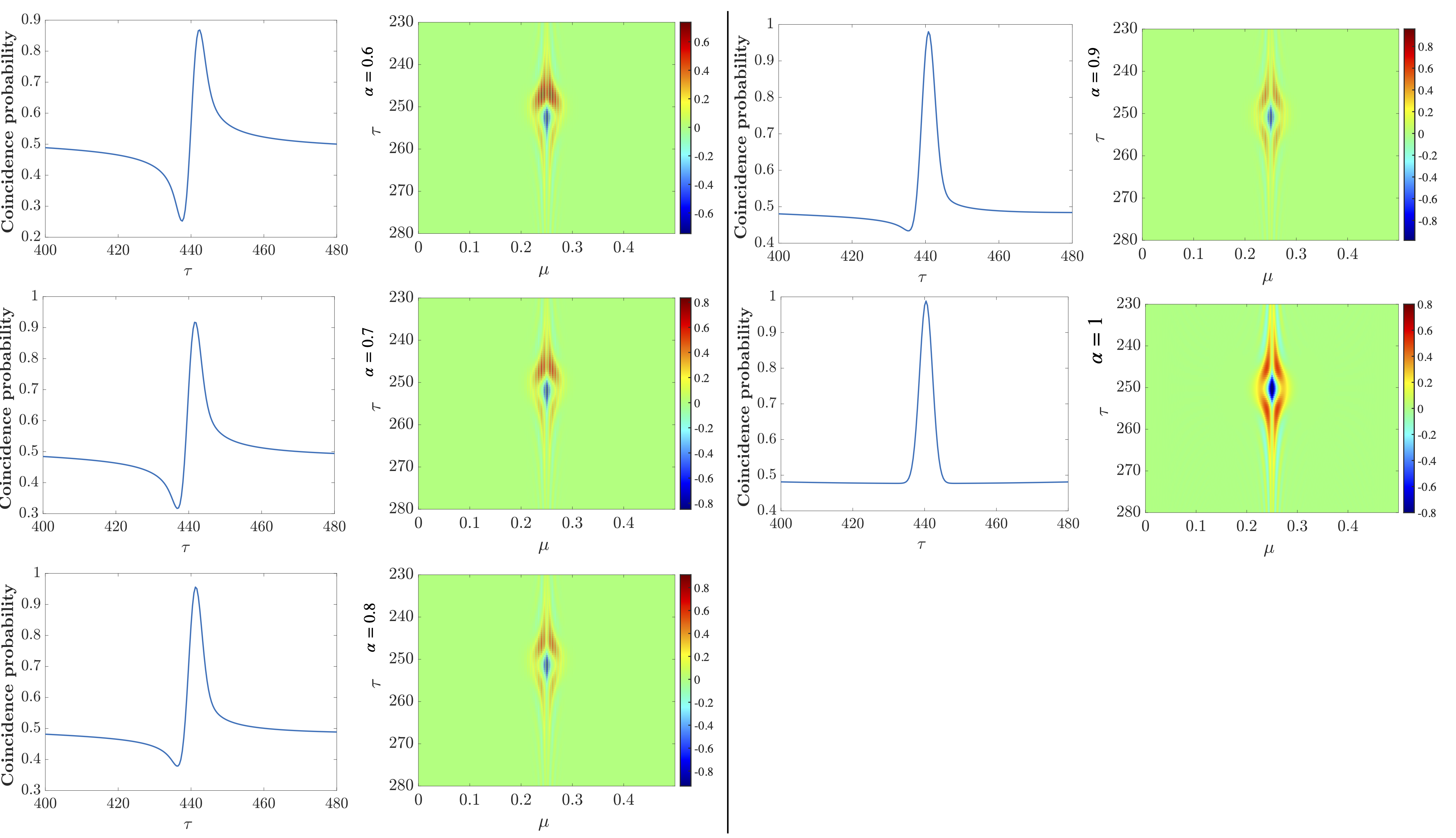}
\caption{\label{an1}Chronocyclic Wigner distribution  of the function Eq.~(\ref{firstcasespec}) and coincidence probability (cut of the Wigner distribution at $\mu=0$) for different values of the anyonic parameter from $\alpha=0$ to $1$ by step of 0.1. The coincidence probability is centered around the zero delay (here centered at the value 440 for numerical reasons). The typical temporal (resp. frequency) unit is the ps (THz) for integrated and bulk photon pair sources and the $\mu$s (resp. MHz) for atomic-cooled sources. For $\alpha=1$ the coincidence probability is a perfect anti-dip without any additional features, but the chronocyclic Wigner distribution reveals an interesting dependence on the frequency.  }
\end{center}
\end{figure*}

\begin{figure*}
 \begin{center}
 \includegraphics[width=1\textwidth]{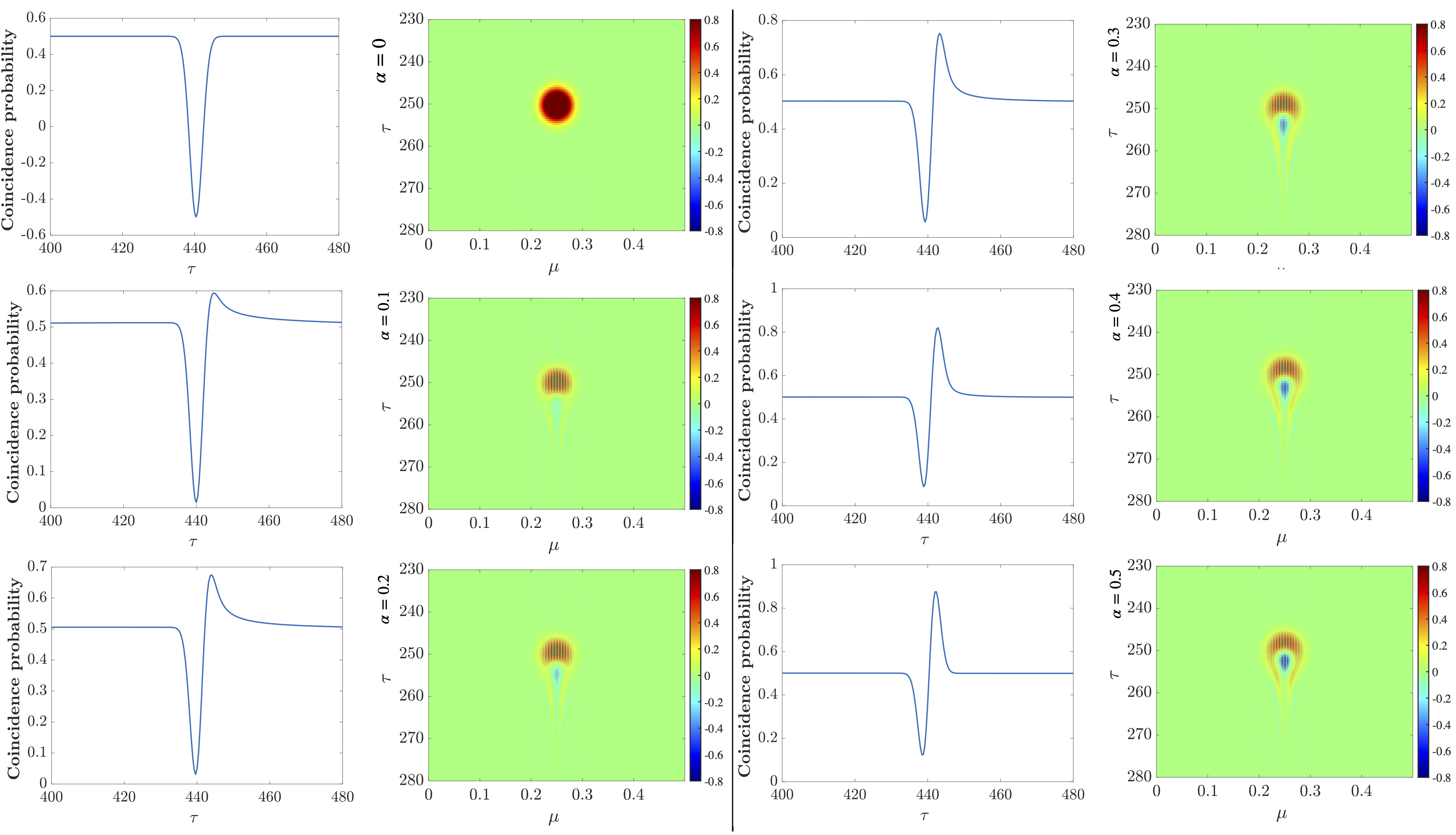}
\includegraphics[width=1\textwidth]{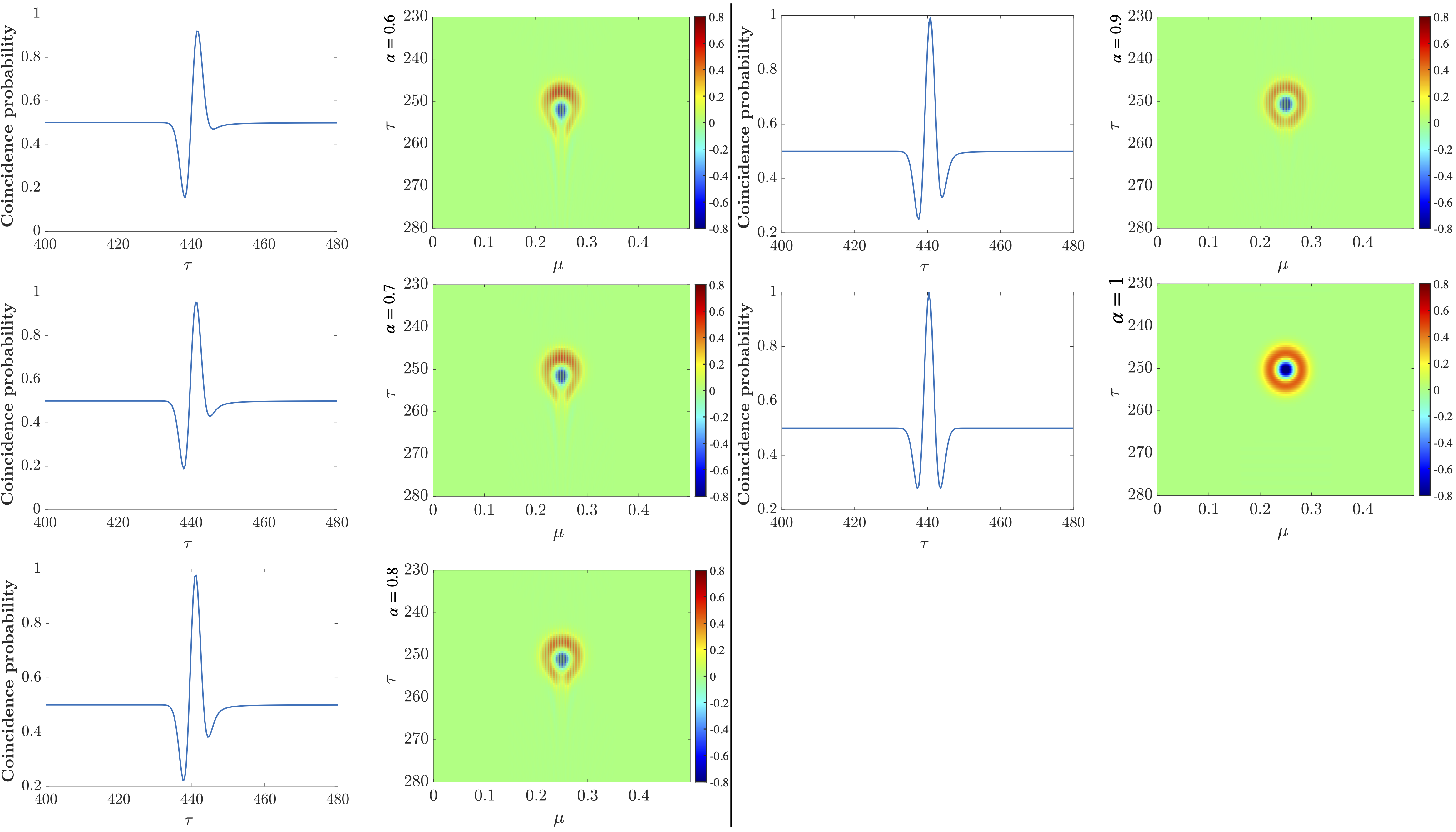}
\caption{\label{an2} Chronocyclic Wigner distribution of the function Eq.~(\ref{anyonsspec}) and coincidence probability (cut of the Wigner distribution at $\mu=0$) for different values of the anyonic parameter from $\alpha=0$ to $1$ by step of 0.1. Shoulders appear for $\alpha=1$ contrary to the previous case.}
\end{center}
\end{figure*}

Particle exchange can be experimentally performed with a beam-splitter, as such an element creates a linear superposition of the wavefunction of the photon pair with its spatially permuted one. The overlap of a photon pair and its associated permutation is measured with the HOM interferometer. The chronocyclic Wigner distribution obtained with generalized HOM interferometry and the associated coincidence probability for $\mu=0$ are represented from $\alpha=0$ to $\alpha=1$ by steps of $0.1$ for the spectral function Eq.~(\ref{firstcasespec}) in Fig.~\ref{an1} and for the spectral function Eq.~(\ref{anyonsspec}) in Fig.~\ref{an2}. Analytical calculation of the coincidence probability as a function of $\tau$ is possible for the spectral function in Eq.~(\ref{anyonsspec}):
\begin{multline}
I(\tau)=\frac{1}{2}(1-\text{cos}(\alpha)\times \sum_{n=0}^{\infty}\frac{1}{n!} \frac{i^{n}}{\sigma^{n+\alpha}} 2^{(\alpha+n-1)}
  \Gamma(\frac{\alpha+n+1}{2})\\
 \cross (1+(-1)^{n+\alpha})\tau^{n})
\end{multline}
where the gamma function is defined as $\Gamma(z)=\int_{0}^{\infty} e^{-t}t^{z-1} dt$ (see \cite{https://doi.org/10.1002/mma.5679} for a proof). We point that the spectral function in Eq.~(\ref{firstcasespec})  has a Gaussian shape for both bosonic $\alpha=0$ and fermionic $\alpha=1$ cases (see Fig.~\ref{an2}). Note that a Gaussian anti-dip shape  was obtained with different degrees of freedom in \cite{walborn_multimode_2003}. However, with the spectral distribution Eq.~(\ref{anyonsspec}), we can observe the presence of shoulders with respect to the central anti-dip in the fermionic case ($\alpha=1$) in the coincidence probability in Fig.~\ref{an2}. Another example of a spectral function that also possesses such shoulders is presented in \cite{francesconi_engineering_2020}. Finally, in Appendix \ref{appendixanyons}, we present another case of anyonic spectrum where the coincidence probability can be calculated analytically.  \\

\subsection{Mach-Zehnder interferometry with photon pairs having an anyonic spectrum}

The coincidence probability for a photon pair with an anyonic spectrum obtained using the MZ interferometer, again considering the factorization given by Eq.~(\ref{normalfacto}), is
\begin{multline}\label{anyonscoinc}
I(\tau)=\frac{1}{2}(1+\text{cos}^{2}(\frac{\alpha}{2})P_{+}(\tau)+\text{sin}^{2}(\frac{\alpha}{2})P_{-}(\tau)).
\end{multline}
We obtain the incoherent sum of the biphoton state with symmetric and antisymmetric spectra with weights $\text{cos}^{2}(\alpha/2)$ and $\text{sin}^{2}(\alpha/2)$. In the bosonic case $\alpha=0$, we recover the Eq.~(\ref{symmetriccase}) and in the fermionic case $\alpha=\pi$ the result in Eq.~(\ref{fermionic}). For $\alpha=0,1$, the full wavefunction is symmetric (resp. antisymmetric). Thus, only the signature of $f_{+}$ (resp. $f_{-}$) is observed, whose oscillation period is determined by the central frequency of $f_{\pm}$.  We point out also that for $\tau=0$, $I(0)=1$ for any $\alpha$. It is not possible to distinguish the bosonic, fermionic and anyonic cases with the MZ interferometer for $\tau=0$ in contrast to the HOM interferometer. Nevertheless, for $\tau\neq 0$, a beating is observed between the central frequencies of the functions modeling the energy conservation and the phase-matching, which reveals the anyonic symmetry.  Note that the intensity signature of anyons with Hanbury-Brown-Twiss interferometer was developed in \cite{PhysRevA.69.063614}.

For the generalized Mach-Zehnder interferometer,  the coincidence probability yields (see Appendix \ref{appendixanyons}) 
\begin{multline}\label{finalanyons}
I(\tau,\mu)=\frac{1}{4}[1+\text{cos}^{2}(\frac{\alpha}{2}) F_{+}(2\tau,\mu)+\text{sin}^{2}(\frac{\alpha}{2}) F_{-}(2\tau,\mu)\\
+\text{sin}(\alpha)\text{Im}((F_{+}(\mu,\tau)+F_{+}(\mu,-\tau))\\
\times (F_{-}(-\mu,-\tau)+F_{-}(\mu,\tau)))].
\end{multline}
In such an expression, the spectrogram intervenes instead of the cosine Fourier transform of the functions $f_{\pm}$ when $\mu=0$. An interference effect between the symmetric and the antisymmetric part is observed through the term $\text{sin}(\alpha)$ which does not cancel owing to the frequency shift. The mathematical reason behind it, is owing to the relation $F^{*}(\mu,\tau)=e^{i\tau\mu}F_{-}(-\tau,-\mu)$. If the frequency shift is set to zero, since $F_{\pm}(\tau)+F_{\pm}(-\tau)=\text{Re}(F_{\pm}(\tau))=P_{\pm}(\tau)$ the interference term thus cancels.

\section{Conclusion}\label{conclusion}
In this paper, we considered the influence of different symmetries of the JSA on the coincidence probability expression measured with the MZ interferometer. We explained the different signatures of symmetric, antisymmetric, and anyonic symmetries obtained with the HOM and MZ interferometers. We discussed a method to vanish the oscillation revealing the coherence of the two-photon state obtained with the MZ interferometer, thanks to a non-linear frequency beam-splitter placed inside the MZ interferometer. We introduced the generalized MZ interferometer consisting of a time and frequency displacement operations performed inside two balanced beam-splitters. The strict equivalent of the experiment described in this paper in current quantum's electronics experiments can be challenging \cite{marquardt_fermionic_2005,chung_quantum_2007} but could be applied with the tools developed in \cite{carrega_anyons_2021}), or by using the valley degree of freedom \cite{jo_quantum_2021}. The frequency shift operation could also be added in other amplitude division-based interferometers, such as the Michelson and Sagnac interferometer.  We also note that recent and promising electro-optics modulators to perform the generalized Mach-Zehnder experiment are presented in \cite{hu_-chip_2021,chen_single-photon_2021}. \\

The development of a simple optical interferometer for sensing of fragile materials was recently developed in \cite{chen_hong-ou-mandel_2019,lyons_attosecond-resolution_2018,ndagano_hong-ou-mandel_2021}. As shown in \cite{scott_noise_2021}, the temporal sensitivity measured with the MZ interferometer is higher than the one obtained with the HOM interferometer, however, it is less robust to photon phase-noise. The perspective of the introduced generalized MZ interferometer could provide a better sensitivity compared to the generalized HOM one \cite{fabre_parameter_2021} for frequency-estimation or multiparameter-estimation protocols.  Finally, different spectral functions of photon pairs can be engineered with a spatial light modulator or by modulating the non-linear susceptibility in bulk non-linear crystal to simulate the statistics of non-Abelian anyons \cite{RevModPhys.80.1083}. Two-photon quantum information protocols with an anyonic symmetry represent another perspective.

\section*{ACKNOWLEDGMENT}
 N.Fabre acknowledges support from the project “Quantum Optical Technologies” carried out within the International Research Agendas programme of the Foundation for Polish Science co-financed by the European Union under the European Regional Development Fund. N.Fabre acknowledges discussions with Perola Milman, Florent Baboux and Sara Ducci.\\
 
\appendix

\section{Expression of the coincidence probability for the MZ interferometer}\label{interferometry}
In this section, we show the expression of the coincidence probability for the MZ interferometer (see Eq.~(\ref{difficultexpression})). Starting from the state defined by Eq.~(\ref{SPDC}), after the first balanced beam-splitter of the MZ interferometer, the wavefunction becomes
\begin{multline}
\ket{\psi}=\frac{1}{2}\int_{\mathds{R}}\int_{\mathds{R}} d\omega_{s} d\omega_{i} \text{JSA}(\omega_{s},\omega_{i}) (\hat{a}^{\dagger}(\omega_{s})+\hat{b}^{\dagger}(\omega_{s}))\\
\cross (\hat{a}^{\dagger}(\omega_{i})-\hat{b}^{\dagger}(\omega_{i}))\ket{0}.
\end{multline}
After a time displacement operation is performed in the spatial mode $a$, the wavefunction becomes
\begin{multline}
\ket{\psi_{\tau}}=\frac{1}{2}\int_{\mathds{R}}\int_{\mathds{R}} d\omega_{s} d\omega_{i}  \text{JSA}(\omega_{s},\omega_{i}) [e^{i(\omega_{s}+\omega_{i})\tau}\hat{a}^{\dagger}(\omega_{s})\hat{a}^{\dagger}(\omega_{i})\\-e^{i\omega_{s}\tau}\hat{a}^{\dagger}(\omega_{s})\hat{b}^{\dagger}(\omega_{i})
+e^{i\omega_{i}\tau}\hat{a}^{\dagger}(\omega_{i})\hat{b}^{\dagger}(\omega_{s})-\hat{b}^{\dagger}(\omega_{s})\hat{b}^{\dagger}(\omega_{i})]\ket{0}.
\end{multline}
After the second balanced beam-splitter, the final wavefunction is:
\begin{multline}
\ket{\psi_{\tau}}=\frac{1}{4}\int_{\mathds{R}}\int_{\mathds{R}} d\omega_{s} d\omega_{i} \text{JSA}(\omega_{s},\omega_{i}) \\
[e^{i(\omega_{s}+\omega_{i})\tau}(\hat{a}^{\dagger}(\omega_{s})+\hat{b}^{\dagger}(\omega_{s}))(\hat{a}^{\dagger}(\omega_{i})+\hat{b}^{\dagger}(\omega_{i}))\\
-e^{i\omega_{s}\tau}(\hat{a}^{\dagger}(\omega_{s})+\hat{b}^{\dagger}(\omega_{s}))(\hat{a}^{\dagger}(\omega_{i})-\hat{b}^{\dagger}(\omega_{i}))\\
+e^{i\omega_{i}\tau}(\hat{a}^{\dagger}(\omega_{s})-\hat{b}^{\dagger}(\omega_{s}))(\hat{a}^{\dagger}(\omega_{i})+\hat{b}^{\dagger}(\omega_{i}))\\
-(\hat{a}^{\dagger}(\omega_{s})-\hat{b}^{\dagger}(\omega_{s}))(\hat{a}^{\dagger}(\omega_{i})-\hat{b}^{\dagger}(\omega_{i}))]\ket{0}.
\end{multline}
We now post-select the part of the wavefunction that leads to coincidences in spatial ports $a$ and $b$ and proceed to a change of variable:
\begin{multline}
\ket{\psi_{\tau}}=\frac{1}{4}\int_{\mathds{R}}\int_{\mathds{R}} (\text{JSA}(\omega_{s},\omega_{i}) +\text{JSA}(\omega_{i},\omega_{s}))(e^{i(\omega_{s}+\omega_{i})\tau}+1) \\
+(e^{i\omega_{s}\tau}+e^{i\omega_{i}\tau})(\text{JSA}(\omega_{s},\omega_{i}) -\text{JSA}(\omega_{i},\omega_{s}))\ket{\omega_{s},\omega_{i}}d\omega_{s}d\omega_{i}.
\end{multline}
If the JSA is symmetric (resp. antisymmetric) the non-frequency-resolved coincidence probability $I(\tau)=\iint d\omega_{s}d\omega_{i} \abs{\bra{\psi_{\tau}}\ket{\omega_{s},\omega_{i}}}^{2}$ can be expressed by Eq.~(\ref{symmetryJSA}) and Eq.~(\ref{fermionic}).

\section{Frequency beam-splitter}\label{Frequencybeamsp}
\begin{figure*}
 \begin{center}
\includegraphics[width=1\textwidth]{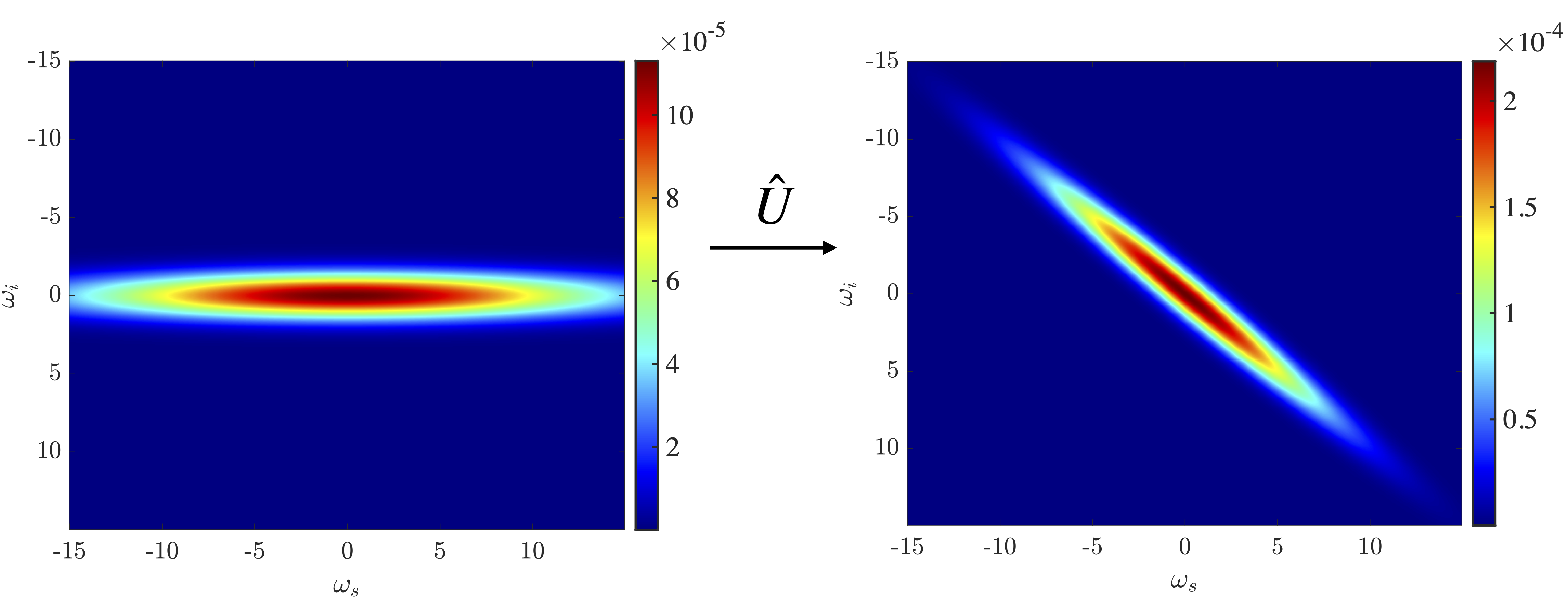}
\caption{\label{JSIFigg}The frequency beam-splitter is a 45 degrees rotation of the JSA, and yields from the presented initial separable state to a spectral entangled state. }
\end{center}
\end{figure*}

In this section, we explain the effect of the frequency beam-splitter operation on an elliptic JSA and emphasize that it is not simply a change of a variable in the integral of Eq.~(\ref{SPDC}). Applying the FBS operation to the separable state in the spectral variables $\omega_{s},\omega_{i}$:
\begin{equation}
\ket{\psi}=\int_{\mathds{R}}\int_{\mathds{R}} f_{+}(\omega_{s})f_{-}(\omega_{i}) \ket{\omega_{s},\omega_{i}}_{ab} d\omega_{s} d\omega_{i}
\end{equation}
we obtain:
\begin{multline}
\hat{U}\ket{\psi}=\int_{\mathds{R}}\int_{\mathds{R}} f_{+}(\omega_{s})f_{-}(\omega_{i}) \ket{\omega_{+},\omega_{-}}_{ab} d\omega_{s} d\omega_{i}\\
= \int_{\mathds{R}}\int_{\mathds{R}} f_{+}(\omega_{+})f_{-}(\omega_{-}) \ket{\omega_{s},\omega_{i}}_{ab} d\omega_{s} d\omega_{i}.
\end{multline}
which is EPR state a spectrally entangled state in the variable $\omega_{s},\omega_{i}$. The operation is a 45 degrees rotation of the JSA in the $(\omega_{s},\omega_{i})$ plane as represented in Fig.~\ref{JSIFigg}.
If the two single photons are in the mode $a$ which can be a polarization or spatial mode, the frequency beam-splitter acts as the identity operator, which is a characteristic of a particular experimental implementation. 

\section{Mach-Zehnder interferometry with fermions}\label{MZfermions}
In this section, starting with a pair of fermions with a symmetric spectrum, we obtain the same expression of the coincidence probability with the MZ as with a photon pair with a antisymmetric spectrum. The fermionic commutation relation relating different creation operators is
\begin{equation}\label{fermionscomm}
\hat{f}^{\dagger}_{a}(\omega)\hat{f}^{\dagger}_{b}(\omega')+\hat{f}^{\dagger}_{b}(\omega')\hat{f}^{\dagger}_{a}(\omega)=0
\end{equation}
where $a,b$ denote here two spatial modes. We consider the initial fermionic state $\ket{\psi}=\iint  F(\omega,\omega')d\omega d\omega' \hat{f}^{\dagger}_{a}(\omega)\hat{f}^{\dagger}_{b}(\omega') \ket{0}$. We suppose that the two (electronic) beam-splitters are balanced, and the reflectivity factor does not depend on the energy. After the second beam-splitter, placing all $\hat{f}^{\dagger}_{a}$ on the left of $\hat{f}^{\dagger}_{b}$ and using Eq.~(\ref{fermionscomm}), we obtain:
\begin{multline}
\ket{\psi_{\tau}}=\frac{1}{4}\int_{\mathds{R}}\int_{\mathds{R}} F(\omega,\omega') [e^{i(\omega+\omega')\tau}(\hat{f}^{\dagger}_{a}(\omega)\hat{f}^{\dagger}_{b}(\omega')-\hat{f}^{\dagger}_{a}(\omega')\hat{f}^{\dagger}_{b}(\omega))\\
+e^{\omega\tau}(\hat{f}^{\dagger}_{a}(\omega)\hat{f}^{\dagger}_{b}(\omega')+\hat{f}^{\dagger}_{a}(\omega')\hat{f}^{\dagger}_{b}(\omega))\\
+e^{i\omega'\tau}(\hat{f}^{\dagger}_{a}(\omega)\hat{f}^{\dagger}_{b}(\omega')+\hat{f}^{\dagger}_{a}(\omega')\hat{f}^{\dagger}_{b}(\omega))\\
+\hat{f}^{\dagger}_{a}(\omega)\hat{f}^{\dagger}_{b}(\omega')-\hat{f}^{\dagger}_{a}(\omega')\hat{f}^{\dagger}_{b}(\omega)]d\omega d\omega'.
\end{multline}
Then, the no-frequency resolved coincidence probability is
\begin{multline}
I(\tau)=\frac{1}{4}\int_{\mathds{R}}\int_{\mathds{R}} | (e^{i(\omega+\omega')\tau}+1)(F(\omega,\omega') -F(\omega',\omega))\\
+(e^{i\omega\tau}+e^{i\omega'\tau})(F(\omega,\omega') +F(\omega',\omega))|^{2}d\omega d\omega'.
\end{multline}
We point out the sign difference between this last expression and the expression obtained with bosons Eq.~(\ref{difficultexpression}) owing to the different commutation relations of creation operators. Then, we find indeed that fermions with a symmetric spectrum give indeed the same the coincidence probability than bosons with an antisymmetric spectrum. Besides, a fermion pair with an antisymmetric spectrum will yield the same coincidence probability as a photon pair with a symmetric spectrum.\\

\section{Coincidence probability expression obtained with the HOM and MZ interferometers for photon pairs with an anyonic spectrum}\label{appendixanyons}

In this section, we calculate the coincidence probability measured using the HOM interferometer for a specific example of the phase-matching function $A(\omega_{s},\omega_{i})=e^{i\alpha \text{sgn}(\omega_{s}-\omega_{i})}$. The coincidence probability is
 \begin{multline}
 I(\tau)=\frac{1}{2}(1-\text{Re}[e^{i\alpha} \int_{-\infty}^{0} d\omega_{-} \abs{f_{-}(\omega_{-})}^{2}e^{i(\omega_{s}-\omega_{i})\tau})\\
 +e^{-i\alpha} \int_{0}^{\infty} d\omega_{-} \abs{f_{-}(\omega_{-})}^{2}e^{i(\omega_{s}-\omega_{i})\tau})],
  \end{multline}
  which can be written as:
   \begin{multline}
 I(\tau)=\frac{1}{2}(1-\text{cos}(\alpha)\times \text{Re}
\int_{0}^{\infty} d\omega_{-} \abs{f_{-}(\omega_{-})}^{2}e^{i\omega_{-}\tau}).
     \end{multline}
 For $\alpha=0,\pi$ we recover the bosonic and fermionic cases respectively. For $\alpha=\pi/2$, the particles become distinguishable owing to an interference effect between the symmetric and the antisymmetric parts of $f_{-}$.

For a photon pair with an anyonic symmetry, the coincidence probability measured with the MZ interferometer is:
\begin{multline}
I(\tau)=\frac{1}{16}\int_{\mathds{R}}\int_{\mathds{R}} d\omega_{s}d\omega_{i} \text{JSI}(\omega_{s},\omega_{i}) |(1+e^{i\alpha})(e^{i\omega_{+}\tau}+1)\\
+(1-e^{i\alpha})(e^{i\omega_{s}\tau}+e^{i\omega_{i}\tau})|^{2}.
\end{multline}
The development of this last expression is composed of self-interference terms which are expressed in Eq.~(\ref{anyonscoinc}) and the interference term is equal to
\begin{equation}
16\ \text{Im}(\text{sin}(\alpha)\text{JSI}(\omega_{s},\omega_{i})P_{+}(\tau/2)P_{-}(\tau/2)).
\end{equation}
Because $\text{JSI}(\omega_{s},\omega_{i})=\abs{\text{JSA}(\omega_{s},\omega_{i})}^{2}$ and the cosine Fourier transform are real quantities, then the interference term is zero and we recover Eq.~(\ref{anyonscoinc}).
For the generalized Mach-Zehnder, the coincidence probability yields
\begin{widetext}
\begin{multline}
I(\tau,\mu)=\frac{1}{16} \int_{\mathds{R}}\int_{\mathds{R}} [\text{cos}^{2}(\frac{\alpha}{2})|\text{JSA}(\omega_{s},\omega_{i})e^{i(\omega_{s}+\omega_{i})\tau}+\text{JSA}(\omega_{s}+\mu,\omega_{i}+\mu)|^{2}\\
 + \text{sin}^{2}(\frac{\alpha}{2})|(\text{JSA}(\omega_{s},\omega_{i}+\mu)e^{i\omega_{s}\tau}+\text{JSA}(\omega_{s}+\mu,\omega_{i})e^{i\omega_{i}\tau})|^{2}\\
 +2\text{cos}(\frac{\alpha}{2})\text{sin}(\frac{\alpha}{2})\text{Im}((|\text{JSA}(\omega_{s},\omega_{i})e^{i(\omega_{s}+\omega_{i})\tau}+\text{JSA}(\omega_{s}+\mu,\omega_{i}+\mu))\\
  \cross (\text{JSA}^{*}(\omega_{s},\omega_{i}+\mu)e^{-i\omega_{s}\tau}+\text{JSA}^{*}(\omega_{s}+\mu,\omega_{i})e^{-i\omega_{i}\tau}))] d\omega_{s}d\omega_{i} .
\end{multline}
\end{widetext}
After considering the factorization Eq.~(\ref{normalfacto}), we obtain Eq.~(\ref{finalanyons}). Using a proof similar to that performed with fermions, we can show that the coincidence probability of a pair of anyons with a symmetric spectrum can be reproduced by a pair of bosons with an anyonic spectrum.

\bibliography{bibliomach}

\end{document}